# High-precision Joint Time and Frequency Transfer over a Fiber-optic Telecom Testbed

Hao Zhang [1], Guiling Wu [1,2], Xinwan Li [1], and Jianping Chen [1,2]
1 State Key Laboratory of Advanced Optical Communication Systems and Networks, Department of Electronic Engineering, Shanghai Jiao Tong University, Shanghai 200240, China
2 Shanghai Key Laboratory of Navigation and Location Based Services, Shanghai 200240, China
E-mail: wuguiling@sjtu.edu.cn

*Abstract*—It is highly desired to operate a joint time and frequency transfer over the exiting fiber-optic telecom networks with the commercial data transmission. Previously, we proposed a time transfer scheme based on bidirectional time division multiplexing transmission over a single fiber with the same wavelength (BTDM-SFSW), and an optical amplification scheme of single-fiber bidirectional-transmission unidirectional optical amplifier (SFBT-UOA). The effects of excess noises like Rayleigh backscattering etc. on the received time signals can be efficiently suppressed for an accepted timing jitter. Simultaneously, the bidirectional propagation delay symmetry can be guaranteed for an expected uncertainty without requiring complicated and expensive link calibration. In this paper, we propose a high-precision joint BTDM-SFSW based time and unidirectional frequency transfer over conventional telecom networks. A stable frequency signal is achieved by compensating the fiber temperature dependent phase variation according to the time transfer. The proposed joint transfer scheme is demonstrated over a 320 km fiber-optic telecom testbed with two 10 Gb/s data transmission channels. The time transfer stabilities (time deviation, TDEV) of 23.2 ps/s and 5.4 ps/$10^5$ s, respectively, and the frequency transfer stabilities (Allan deviation, ADEV) of 5.4×$10^{-13}$/s and 9.5×$10^{-17}$/$10^5$ s, respectively, are reached. The measured bit error rates (BER) of the 10 Gb/s communication data are well below $10^{-13}$.

## I. INTRODUCTION

The needs for high-precision time and frequency standards in a variety of applications, such as time and frequency metrology, navigation and astronomy etc., have immensely promoted the development of fiber-optic time and/or frequency transfer [1]. Considering the complexity and cost, it is desirable to implement time and/or frequency transfer over the existing wide-spread fiber-optic telecom networks. However, different from the fiber-optic telecom networks adopting single-fiber unidirectional transmission, almost all the high-precision fiber-optic time and/or frequency transfer schemes [2-7] adopt single-fiber bidirectional transmission to deduct the influences from fiber link propagation delay and its variation.

Previously, we proposed a time transfer scheme based on bidirectional time division multiplexing transmission over a single fiber with the same wavelength (BTDM-SFSW) [8], and an optical amplification scheme of single-fiber bidirectional-transmission unidirectional optical amplifier (SFBT-UOA) [9]. The effects of excess noises like Rayleigh backscattering etc. on the received time signals can be efficiently suppressed for an accepted timing jitter. Simultaneously, the bidirectional propagation delay symmetry can be guaranteed for an expected uncertainty without the cost of complicated and expensive link calibration. Moreover, the proposed SFBT-UOA also has the potential to support both bidirectional time transfer and unidirectional conventional network services at the same time. In this paper, we propose a high-precision joint time and frequency transfer over conventional telecom networks based on BTDM-SFSW time transfer and unidirectional frequency transfer. The stable fiber-optic frequency transfer is achieved by compensating the fiber link propagation delay variation based on the time transfer. A 320 km fiber-optic telecom testbed is established with the commercial 10 Gb/s data transmission, which verifies the feasibility of the proposed scheme and its capacity over the conventional fiber telecom networks to the most degree.

## II. SYSTEM CONFIGURATION

Fig 1 illustrates the established fiber-optic telecom testbed. For time transfer, the timing signal (e.g. one pulse per second, 1 PPS) from a common Rb clock (Symmetricom, 8040C) is provided for both sites to eliminate the effect of clock drift on the test. The 1 PPS at site A is encoded into a time code by a dedicated encoder with a low delay variation, similar to the method presented in [10]. The delay of 1 PPS at site B is completed by a time delay adjuster (TDA). Then the delayed 1PPS is encoded into a time code as well. The generated time codes are modulated on the optical carriers through commercially-available SFP transceivers, which have an output power of 0 dBm, a central wavelength $\lambda_1$ of 1549.32 nm and a linewidth of 20 pm. By the control of the corresponding optical switch (OS) with a switching time of about 1 ms in maximum, the optical signals at both sites are launched into the fiber link only during the transmission of the time codes. The 1 PPS in each receiving time code is extracted by the decoder at the corresponding site. The time interval between the local input 1 PPS and the received one is determined by the time interval counter (TIC, Stanford Research System, SR620) at each site.

The delay of the TDA, $T_d$, at site B is also measured by a SR620. The clock difference can be calculated as follows,

$$\Delta T = \frac{1}{2}\left[(T_{AB} - T_{BA} - T_d) + (\tau_{AB}^F - \tau_{BA}^F) + (\tau_A^T - \tau_A^R + \tau_B^R - \tau_B^T)\right] \quad (1)$$

where $T_{AB}$ ($T_{BA}$) is the measured time difference between the local time signal and the received one at site A (B), $\tau_{AB}^F$ ($\tau_{BA}^F$) is the propagation delay of fiber link from site A to site B (site B to site A), $\tau_A^T$ ($\tau_B^T$) and $\tau_A^R$ ($\tau_B^R$) are the sending and receiving delays at site A (site B), respectively.

For frequency transfer, a radio frequency signal of 1 GHz is generated and referenced to the 8040C. With the modulation of a SFP transceiver, the frequency signal passes through the fiber link and reaches site B over the wavelength $\lambda_2$=1548.52 nm. At site B, the receiving frequency signal is converted by another SFP transceiver and filtered out by an electrical bandpass filter due to the digital signal process in the SFP transceivers. Passing through a phase shifter (PS) controlled according to the measured fiber link propagation delay variation, $\Delta\tau^F$, by time transfer (see (2)), a stable sinusoidal frequency output can be achieved at site B. In the experiment, the measured $\Delta\tau^F$ is processed with a 200 moving average due to the uncertainty of SR620. Moreover, the PS is an ideal one and will be implemented by hardware in the next work.

$$\Delta\tau^F = \frac{1}{2}\Delta(T_{AB} + T_{BA} - T_d) \quad (2)$$

For the conventional communication services, two optical carriers from the SFP transceivers with the wavelengths of $\lambda_3$=1550.12 nm and $\lambda_4$=1550.92 nm are modulated at 10 Gb/s in a non-return-to-zero (NRZ) modulation format with a $2^7-1$ pseudorandom binary sequence (PRBS). After a 170 km and 320 km transmission, respectively, the optical carriers are received by the transceivers and sent into the bit error rate tester (BERT) for BER evaluation.

As shown in Fig 1, site A and site B are connected by a 320 km G.652 fiber, the corresponding dispersion compensated fiber (DCF) and four SFBT-UOAs. In the SFBT-UOA, a commercially-available EDFA with optical isolators, four WDMs with a channel spacing of 0.8 nm, and a 2×2 OS with a switching time of 1 ms in maximum are employed. The three WDMs with a channel spacing of 0.8 nm in Fig 1 are employed to separate and combine the optical signals of time transfer, frequency transfer and conventional network services. It should be noted that the link access control mechanism [9] is required to implement a BTDM-SFSW based time transfer employing SFBT-UOAs.

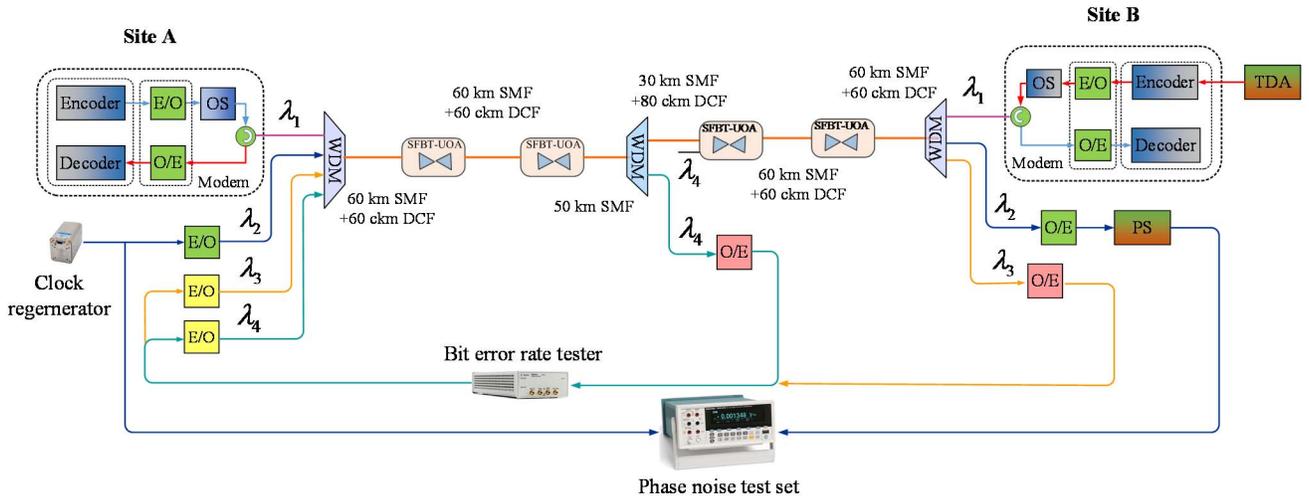

Fig 1 Experimental testbed of a high-precision joint time and frequency transfer together with 10 Gb/s data transmission. TDA: time delay adjuster; PS: phase shifter; BERT: bit error rate tester.

### III. EXPERIMENTAL RESULTS AND DISCUSSIONS

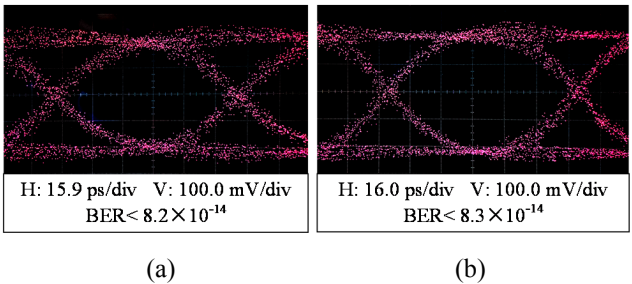

Fig 2 Measured BER for the 10 Gb/s data with the 170 km (a) and 320 km (b) transmission.

The measured BER for the 10 Gb/s data with a 170 km and a 320 km transmission, respectively, are given in Fig 2. From the figure, the measured eye diagram can be clearly recognized. In the duration of more than 20 min, no bit error is observed, which means that the BER is lower than $8.2\times10^{-14}$ and $8.3\times10^{-14}$, respectively. The obtained results suggest that the SFBT-UOA has the capacity to support bidirectional time transfer, unidirectional frequency transfer and conventional network services at the same time. And the service quality is sufficient for conventional data communication.

Fig 3 (a) illustrates the stabilities (time deviation, TDEV) of time transfer over a 2 m and a 320 km fiber link. The TDEVs over the 2 m fiber link amounts to 19.6 ps/s and 1.8 ps /$10^5$s. And the adopted codecs, SFP transceivers and the TICs accounts for the substantial part since the impact of such a short fiber is negligible. As the link extends to 320 km, the TDEVs are deteriorated to 23.2 ps/s and 5.4 ps/$10^5$s. The aggravation of the short-term stability [11] mainly comes from the degradation of signal-to-noise ratio (SNR) against the increase of fiber length. Nevertheless, the degradation of the long-term stability [11] is mostly attributed to the increased fluctuation of the bidirectional propagation delay asymmetry, resulted from the variations of fiber temperature and the transmitting wavelength difference.

Also plotted in Fig 3 (a) are the stabilities (Allan deviation, ADEV) of 1 GHz frequency transfer over a 2 m and a 320 km fiber link. The ADEVs of $8.6 \times 10^{-15}$/s and $4.6 \times 10^{-18}$/$10^5$ s can be achieved over the 2 m fiber link. It can be regarded as the stability floor for fiber-optic frequency transfer, which is dominated by the SFP transceivers and the electrical devices (e.g. low noise amplifiers, mixers) adopted for signal conversion. As the link length reaches 320 km, the phase of the frequency signal varies significantly with a peak-to-peak value of about 19 ns (see Fig 3 (b)) due to the fiber temperature variation. Thanks to the cooperation of the ideal PS, a stable frequency signal can be received at site B with a peak-to-peak phase variation of less than 30 ps, which manifests the ADEVs of $5.4 \times 10^{-13}$/s and $9.5 \times 10^{-17}$/$10^5$ s. For a better stability performance, the receiving SNR and the uncertainty of the deployed TICs can be improved for short averaging time while the fiber link propagation delay stabilization, such as external temperature control of SFP transceivers [12], temperature-dependent wavelength dispersion calibration etc. can be employed for long averaging time.

## IV. CONCULSION

In summary, we proposed a high-precision joint time and frequency transfer over conventional telecom fiber networks, and successfully demonstrated a BTDM-SFSW based time transfer and unidirectional frequency transfer over a 320 km fiber-optic telecom testbed in the laboratory environment. All the devices employed are commercially available and compatible with the conventional telecom networks. The time transfer stabilities of 23.2 ps/s and 5.4 ps/$10^5$ s, respectively, and the frequency transfer stabilities of $5.4 \times 10^{-13}$ and $9.5 \times 10^{-17}$, respectively, are demonstrated. The measured BER for the 10 Gb/s communication data is well below $10^{-13}$. The proposed scheme validate the capacity of conventional telecom networks to support time transfer, frequency transfer and communication services at the same time, and thus provides possibility to embed a high-precision time and frequency transfer into long-haul commercial fiber transmission links.

## ACKNOWLEDGMENT

This work was supported in part by the National Natural Science Foundation of China (61627817, 61535006).

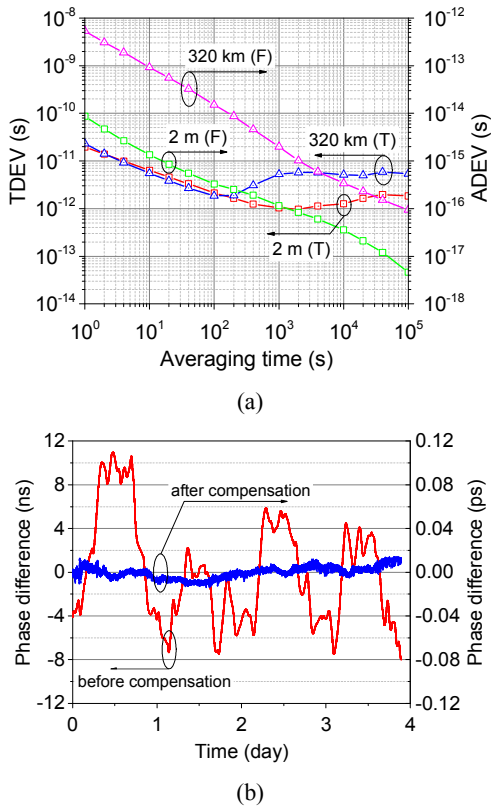

Fig 3 (a) time deviation (TDEV) of time transfer and Allan deviation (ADEV) of frequency transfer over a 2 m and 320 km fiber link; (b) phase variation before and after phase shifter (PS) over the 320 km fiber link.